\documentclass[format=sigconf]{acmart}
\usepackage[T1]{fontenc}
\usepackage{paralist}
\usepackage{amsmath}
\usepackage{geometry}
\usepackage{graphicx}
\usepackage[inline]{enumitem}
\usepackage{setspace}
\usepackage{fancyhdr}\usepackage{here}
\usepackage{wrapfig,subfigure}

\begin{document}
\newcommand{\tx}{Having your cake and eating it too: Scripted
  workflows for 
  image manipulation}
\title{\tx}
\author{Paul A. Thompson, Ph.D}
\affiliation{\institution{Sanford Research \& Health}}
\affiliation{\institution{University of South
    Dakota-Vermillion}
}
\email{paul.thompson@sanfordhealth.org}
\author{Norm Matloff, Ph.D.}
\affiliation{\institution{University of
    California-Davis}
}
\email{nsmatloff@ucdavis.edu}
\author{Alex Fu}
\affiliation{\institution{Princeton University}}
\email{aafu@ucdavis.edu}
\author{Ariel Shin}
\affiliation{\institution{University of
    California-Davis}
}
\email{arishin@ucdavis.edu}

\begin{abstract}
  The reproducibility issue in science has come under increased
  scrutiny. One consistent suggestion lies in the use of scripted
  methods or workflows for data analysis. Image analysis is one area
  in science in which little can be done in scripted methods.  The
  SWIIM Project (\textbf{S}cripted \textbf{W}orkflows to
  \textbf{I}mprove \textbf{I}mage \textbf{M}anipulation) is designed
  to generate workflows from popular image manipulation tools. In the
  project, 2 approaches are being taken to construct workflows in the
  image analysis area. First, the open-source tool GIMP is being
  enhanced to produce an active log (which can be run on a stand-alone
  basis to perform the same manipulation). Second, the R system Shiny
  tool is being used to construct a graphical user interface (GUI)
  which works with EBImage code to modify images, and to
  produce an active log which can perform the same operations. This
  process has been successful to date, but is not complete. The basic
  method for each component is discussed, and example code is shown.
\end{abstract}
\maketitle
\renewcommand{\shortauthors}{Thompson et al}
\footnote{The authors would like the acknowledge the contributions of
  Douglas Cromey, MS, University of Arizona, to the ideas in the SWIIM
  project.} 

\section{Introduction}
\subsection{Workflows in science}
A scientific workflow is a tool to structure and regularize a process
(\cite{Gil-2007-1}
\cite{Lushbough2011} %
\cite{Workflow-Model-1995}).
A good description is as follows:
\begin{quote}
  ``Scientific workflows attempt to automate repetitive computation
  and analysis by chaining together related processes. Automating
  repetitive time-consuming tasks allows scientists to keep pace with
  ever-growing volumes of data. Furthermore, workflows can aid in the
  reproducibility of scientific computations by providing a formal
  declaration of an analysis. Reproducibility is central to the
  scientific method, and detailed workflow provenance in- formation
  ensures an analysis can be reproduced and extended.''
  \cite{Lushbough2011} %
\end{quote}
Workflows have been the subject of much investigation.  They take many
different forms. Some workflows are defined by an interactive process,
while others are defined by scripts. In using a scripted workflow, the
process which the workflow performs becomes public, transparent, and
reproducible. 

\subsection{Reproducibility}
Reproducible research methods are
increasingly important in science%
(\cite{Gandrud-2013}%
\cite{Ioannidis-2005-696-701}%
\cite{Ioannidis-2014-e1001747}%
\cite{Maxwell-2012-487-498}%
\cite{Peng-2006-783-789}%
\cite{Peng-2011-1226-1227}%
\cite{Stodden-2014-2}%
\cite{Thompson-2011-11}).
``Reproducibility of research'' is defined by the different issues
which result in problems obtaining the same results from a study. Are
the results the same from a second processing of the same data?

``Getting the same result'' can mean different things. The most
specific is seemingly the simplest: When reanalysing a given dataset
using the same methods, can identical outcome values (test statistics,
p values, summary statistics) be obtained? This may be termed ``data
reproducibility''.  A somewhat different form  of reproducibility may
involve running the study again with different subjects to examine the
``scientific reproducibility'' of the study. Each type of
reproducibility examines the different aspects to the degree to which
the results of a study are repeatable. Science involves determining a
process which can be repeated and produce the same results, and thus
reproducibility is the essence of science.

Obtaining the same result from a given set of data sounds obvious and
trivial, but there are a number of reasons why this can be
problematic. First, certain types of analyses are not closed-form but
rather are iterative and approximating (with a loss function and
convergence criteria). Unless the same convergence criteria, start
values, and step sizes are used, it is entirely possible to get
different outcomes. This is particularly true in cases in which the
outcome surface is relatively flat. Second, analyses may be done in an
interactive manner, and thus the tracking of the exact processes
involved can sometimes be difficult. When interactive methods are
used, it is possible that steps are forgotten, or that steps are done
in different orders, or that the specific details in a step are not
correctly noted. Third, the version of software used for analysis
may change from one use to the next. Newer versions can include
different convergence criteria or even different methods for
estimation. Fourth, the persons using the software can be different,
and use the software in different ways. In the well-known Potti et al
case, the original data were analyzed by a physician who was not well
trained in proper data analysis, proper data storage, or proper use of
training and validation samples (\cite{Baggerly-2011-1}
\cite{Reich-Nature-2011}). Later analysis by better-trained
bioinformatics scientists found many errors, including changes in the
version of the main data analysis tool (\cite{Baggerly-2011-1}).

\newcommand{\holda}{
In the Potti case, other errors included questions about the integrity
of the sample \cite{Baggerly-2011-1}. In many studies, the sample
being examined changes, and unless care is taken to ``freeze'' the
sample, different analyses may be made from different samples.\par
Many of the problems in reproducibility
of results occur because the analysis is incompletely documented or
tracked. Analysis of data begins with data management. Data are
obtained from sources (data collection tools). The data are filtered
(i.e., invalid values are removed or corrected, incorrect cases are
sometimes removed if they were incorrectly added). Data management is
a key step in the process and must be carefully documented for later
checking and examination.
}

In producing scientific articles, the data must be structured for the
analysis first. This is the ``data management'' process, and is often a key
step in the process. Values are corrected. Occasionally data are
removed.
The statistical analysis which examines the data is next
performed. Again, this must be carefully documented to produce valid
outcomes
(\cite{Peng-2011-1226-1227} %
\cite{Donoho-2010-385-388} %
\cite{Sandve-2013-e1003285} %
\cite{Vandewalle-2009-37-47}). %
Scripted methods (i.e.,
analysis performed using programs of computer code) are necessary for
reproducible results
(\cite{Peng-2006-783-789} %
\cite{Donoho-2010-385-388} %
\cite{Sandve-2013-e1003285} %
\cite{Peng-2011}).
The code can be inspected, transfered to others,
used on more than one project, and modified easily. It also functions
as the memory of the project%
(\cite{Thompson-2016-2}).

\newcommand{\holdb}{ Analysis code has another key feature. It is
  ``transparent'' or able to be inspected by others. Rather than
  analyses performed in a ``black box'' of proprietary secret methods,
  scientific data analysis must be performed in an open and clear
  manner. Scripted methods are the key here. While computer programs
  have coded values that require training to use and understand, any
  user with this training can follow the analysis, with the assistance
  of ``comments'' (statements of explanation which should be inserted
  into the code to document the process).  }

The use of analysis code also is ``transparent'' or able to be
inspected by others. Transparent, scripted code ensures that the
author of a scientific document can produce the same results later,
and can demonstrate to others (e.g., journal editors, colleagues)
exactly how the published information was created from source
materials. In science, repeating an analysis must produce the same
result.%

\subsection{Image manipulation}
Scientific image manipulation is a key part of many areas,
particularly basic biology and chemistry
(\cite{CouncilofScienceEditors-2016} %
\cite{Cromey-2010-639-667} %
\cite{Nature-Editors-2016} %
\cite{Newman-2013} %
\cite{Rossner-2002-1151} %
\cite{Rossner-2004-11-15} %
\cite{Rossner-2006-24-25}). %
It is the process of preparing images for publication.  Scientific
journals have clear and well-defined requirements for proper
preparation of images (\cite{CouncilofScienceEditors-2016} %
\cite{Nature-Editors-2016} %
\cite{Rossner-2004-11-15}).  Such scientific image manipulation
follows general guidelines:
\begin{enumerate}[label=(\arabic*)]
\item specific features may not be changed or modified;
\item adjustments to the full image (brightness, contrast color
  values)  are usually
  acceptable;
\item if separate images are grouped together, this must be explicit; and
\item the original image must be retained and be available for examination.
\cite{CouncilofScienceEditors-2016}
\end{enumerate}
There is a clear and well defined difference between preparation of
scientific images and preparation of {\ae}sthetic images
(\cite{NCBEditorial-2006-101} %
\cite{NCBEditorial-2006-203} %
\cite{Couzin-2006-1866-1888}).
Methods acceptable for {\ae}sthetic image preparation include many
techniques which would not be allowed in scientific images.

Image processing is primarily done using interactive tools such as
Adobe Photoshop%
(\cite{Adobe-Photoshop})
ImageJ,%
(\cite{ImageJ-UG-2012})
and GIMP.%
(\cite{GIMP-Documentation-Team-2016})
These programs can read in images, modify them in many ways, and save
the results.  It is sometimes difficult to reproduce the interactive
process of producing an image for a publication from a source
image. This is due in part to the use of the computer mouse, and
partly due to the difficulty of remembering operations.

When images are prepared for scientific presentation, reproducibility
problems are common. The difficulties in reproducibility, due to the
interactive nature of the process, partly arise due to the
``semi-continuous'' nature of the process. When cropping (selecting a
small part of the picture for presentation), a selection is made using
the mouse. Although this is done using positions which are numbers,
the scale is large and the position is difficult to remember
exactly. When increasing brightness-contrast, the increases are done
using a scale which emphasizes relative amounts; the exact value is a
number, but the number is likely not remembered exactly. While a
person could remember such values, the exact numbers are quite
difficult to remember, and the process is not condusive to simple
recollection. 

Image fraud is a serious and pressing issue in
science%
(\cite{Cromey-2010-639-667} %
\cite{Newman-2013} %
\cite{Rossner-2006-24-25} %
\cite{Parrish-2009-161-167}). %
Image fraud includes a number of processes (e.g., image reuse,
improper preparation, improper combination of images). The
``Retraction Watch'' blog provides a contemporaneous record of
research fraud.%
\cite{Marcus-Oransky-RW} %
In examining this blog, it is clear that a large proportion of
retractions involve image fraud. As of 2017/03/31, 512 of the entries
in the blog are related to image fraud.  Improper
image preparation occurs commonly; some reports suggest that 25\% of
all submissions to journals have improper image preparation.%
\cite{Rossner-2006-24-25} %
20 years of discussing the problem have not reduced the incidence of
the problem. Different approaches are needed.

\subsection{Journaling}
Writing code for analysis is a difficult skill. Interactive methods
for data analysis are preferred by some as being simpler and more
intuitive. When an analysis is performed by a graphical user interface
(GUI; a window with buttons and controls), this is termed an
``interactive approach''.  The reproducibility of interactive
approaches is questioned by many.%
\cite{Sandve-2013-e1003285} %
That is because interactive methods often involve important but small
decisions, which are often hard to remember and write down. Details
are difficult to remember correctly.
If the interactive process is at all involved or complicated, the many
decisions which are made ``on the fly'' are difficult to remember
later, and may be hard to communicate in a scientifically complete
manner.

There is a middle ground. In many high-level programs (e.g., SAS/JMP%
\cite{JMP-SAS}%
), the analyst can perform the analysis interactively, while the program
simultaneously creates code which performs the same analysis. This
process of program-created code is termed ``journaling''. The
journaling process creates a log or record, which can then be used to
perform the analysis again. This is available in some but not all
interactive analysis programs. It is available in R using the
``analysis history'' tool.%
\cite{The-R-Foundation}

A ``journaling'' approach to image processing is needed. In the SWIIM
project, several approaches to journaling are being implemented.  The
project is creating such an image-manipulation journaling tool by
working with open-source tools, to produce an executable log (the
journal) of the analysis performed using the tool. The GIMP program
will be enhanced to perform a journaling function, by adding code to
the program. With the R system, a set of tools will be added to the
existing methods to perform programmatic image analysis. In both
cases, the modifications will allow the user
\begin{enumerate}
\item to examine what specifically was done with the image;
\item to perform the same modifications when ``replayed'' on the
  original image; and
\item to step through the modifications to examine them in detail.
\end{enumerate}

Fraudulent and inappropriate image manipulation is a serious problem
in science.  The use of interactive methods makes it very difficult to
provide a clear tracking of all processes performed on an image. A
methodology which performs image manipulation using scripting can
provide transparency in processing, but this is difficult to learn and
use. The best approach to improve reproducibility of image
manipulation is a journaling approach where valid operations performed
interactively would produce code. The code could be run to produce the
same result, and the code could be examined to see what had been done.

\section{Methods}
The SWIIM project is designed to
generate workflows from popular image manipulation tools. Two basic
approaches are used in this effort:
\begin{enumerate}
\item GIMP: modify an existing open-source image manipulation tool to
  journal, or produce an active log which can manipulate images.
\item R: use open-source tools in the R system to manipulate images in
  a GUI.
\end{enumerate}

Contemporary image manipulation toolkits include a large number of
functions, many of which are strictly {\ae}sthetic, and inappropriate
for the preparation of scientific images. The SWIIM project will
concentrate on the journaling process for the following operations:
\begin{enumerate}
\item Import files with formats of jpg, tiff, png, bmp
\item Rotate image 90, 180, 270
\item Flip image through 3rd dimension (vertical, horizontal)
\item Crop image
\item Brightness-contrast adjustment on image
\item Color balance adjustment on image
\item Threshold adjustment and histogram balance
\item Meld images (insert one image into another) with borders
\item Hue (red, blue, green) adjustment
\item Export file with formats of jpg, tiff, png, bmp
\end{enumerate}

\begin{figure*}
\includegraphics[width=\textwidth]{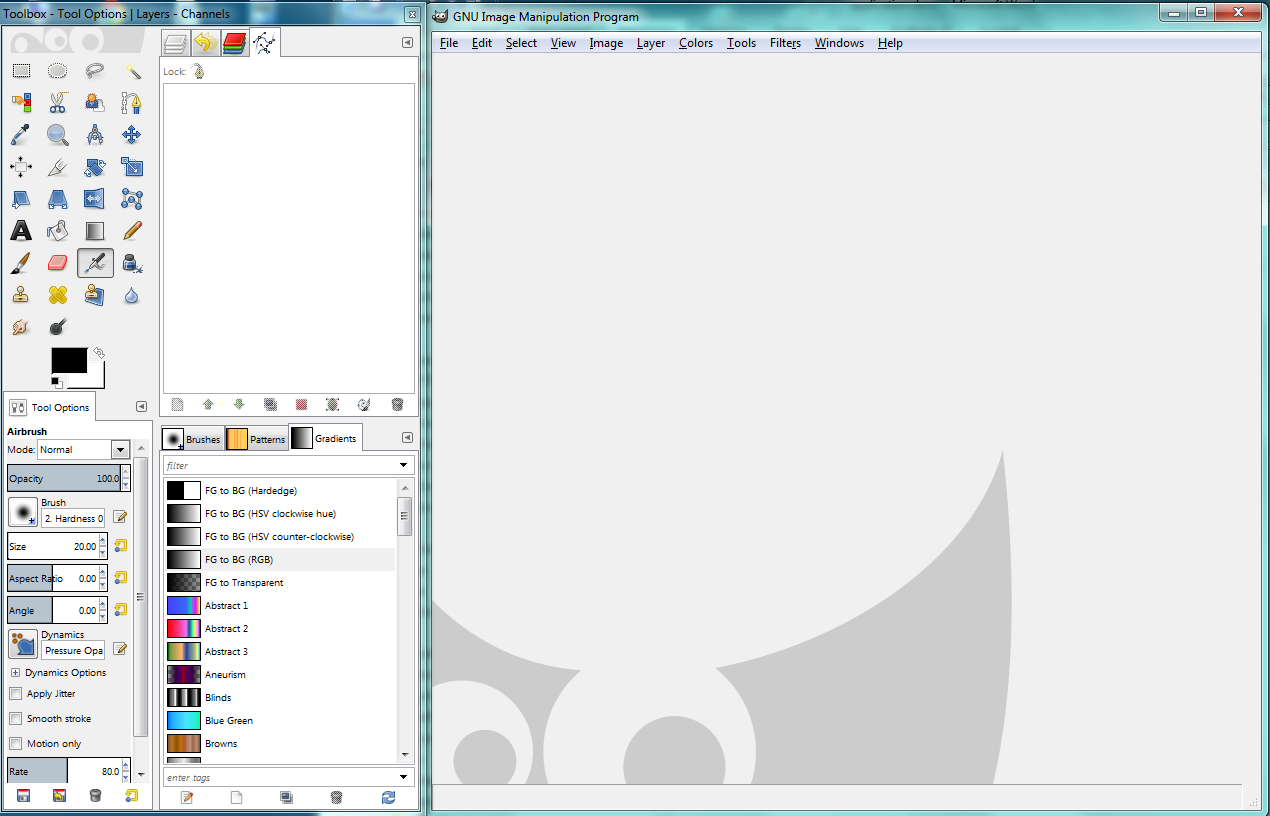}
\caption{GIMP Window Setup}
\label{fig:GIMPwin}
\end{figure*}

\subsection{Application: GIMP}
GIMP is a full-featured image manipulation and modification program
which can perform many types of image manipulations, including all
scientific image manipulations.%
\cite{GIMP-Documentation-Team-2016} GIMP processing involves using a
 GUI which allows
images to be imported, manipulated, modified, and altered, and then
saved in a variety of output formats. The functions for image
manipulation can be invoked by clicking buttons, selecting items from
pull-down lists, or typing keys. In addition, a full-featured
scripting language (script-fu) can be used to construct scripted
programs for image manipulation.

GIMP is written in C.%
\cite{GIMP-Documentation-Team-2016} %
A journaling function will be added by modifying GIMP source code 
(\cite{Budig-GIMP-Script-FU} %
\cite{Peck-2017} %
\cite{Peck-Scripts-2016}) or by creating a ``plug-in''.
In GIMP, the ``parasite'' system is being used to retain the sequence
of actions which have been applied to a given image on the way from
input to export. The parasite system allows the user to associate an
arbitrary string to an image or processing session. At a specific
point in the processing, the information in the ``parasite'' system
can be recovered and used to create the journal for the process, and
this journal can be saved to an output file. The process involves
determining the exact actions which occurred, as well as determining
the values of indexes or locations used in the process. This requires
determining where key strokes and button selections (which had led to
a specific modification point) are processed. Additionally, the values of
selections (i.e., sliders, verniers) must recovered.  Creating this
kind of code-generating add-on or addition to GIMP is an objective of
the GIMP community, and has support from developers and maintainers.

A code system is needed to provide  the journaling function for the
interactive manipulations. There are two which are being set up in
GIMP. These are basically 2 code systems which will be used for the
GIMP component. First, the
 ImageMagick system, a function-base graphics   toolset%
\cite{ImageMagick-Studio-2016} will be used. ImageMagick has
well-defined functions which perform each of the actions defined
above. For a more faithful and complete emulation of code to the
interactive process, the  script-fu system will be employed. This is a
stand-alone GIMP language which is
used to write scripts%
\cite{Good-Python-GIMP}%
\cite{Henstridge-Python-GIMP}
ImageMagick is somewhat less complicated and difficult than is
script-fu, which is a fully-featured programming language. By using
ImageMagick code, simple and equivalent modifications can be made to
images. The advantage of this approach is that ImageMagick can be used
to do a proof of concept, which will be followed up by script-fu over time.
While ImageMagick can perform functions which are equivalent to those
in GIMP, script-fu will perform functions which are exactly the
same. Using script-fu functions will produce an exact match of
images. Using ImageMagick will produce a visually equivalent image,
but the image will not match on the pixel level. 

\subsection{Application: R/Shiny}
The R system is a general-purpose system of statistical, data
management, and data display techniques. It is open-source, and is
produced by a number of different contributors
(\cite{The-R-Foundation}). %

The R system has the Shiny GUI building application
(\cite{Shiny-R-Homepage}). %
Using Shiny, a GUI can be built with buttons, controls, and menus to
control processes. The processes controlled  here are image
analysis and image manipulation processes. The tool which is being
constructed is called ShinyImage.  For image manipulation,
the ShinyImage tool uses the EBImage toolkit%
(\cite{EBImage-2017}) %
for image processing. Performing operations
involves using the GUI to select an operation, executing that
operation on the image, and reflecting the change in the image in the
Shiny GUI.  The journaling process involves writing the commands to a
journaling log file. The log file can be examined for a record of the
processing, and executed to perform the operations again.

\begin{figure*}[t]
\includegraphics[width=\textwidth]{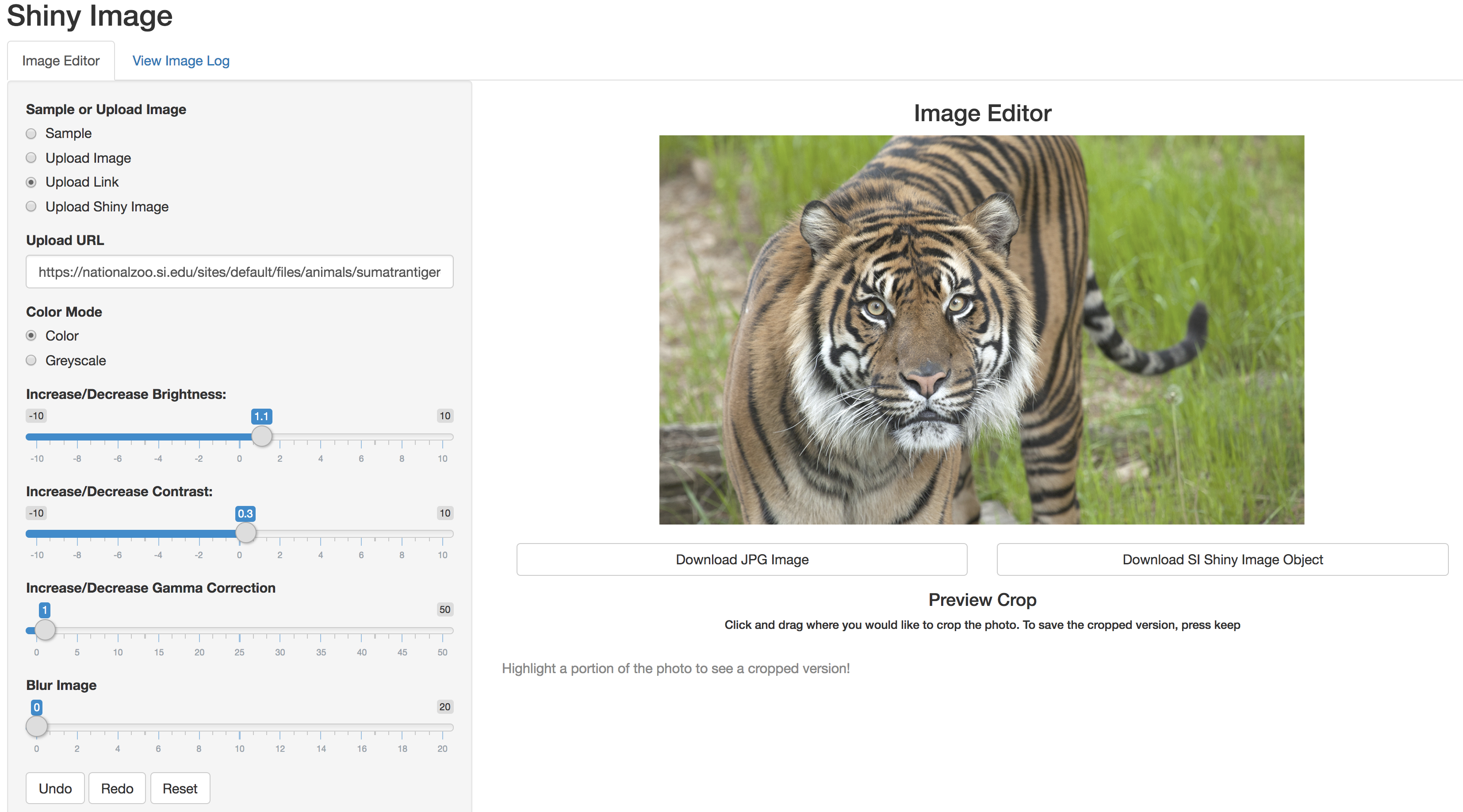}
\caption{ShinyImage GUI}
\label{fig:shinyimage}
\end{figure*}

\section{Status of project}
The SWIIM project is coming to the end of the first year. Progress
been made, but the modifications are not complete.

\subsection{Developments to date: GIMP}
\begin{figure}[H]
\includegraphics[width=.75\columnwidth]{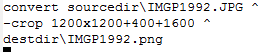}
\caption{ImageMagick code}
\label{fig:IMcode}
\end{figure}

\begin{enumerate}
\item Full compile has been achieved. This requires that all
  supporting programs and files be obtained.
\item The location of ``open file'' operation has been
  identified. File name has been written to an ImageMagick file. The
  separate file can be  processed to modify the image.
\item The location of the ``crop'' operation has been identified and
  this has been translated into ImageMagick code, and written to a
  file.
\item The location of the ``brightness-contrast'' operation has been
  identified and this has been translated into ImageMagick code, and
  written to a file. The different systems use different values for
  the brightness and contrast modifications, but this has been
  correctly aligned.
\item The location of the ``hue'' modification operation has been
  identified and this has been translated into ImageMagick code, and
  written to a file. The different systems use different values for
  the brightness and contrast modifications, but this has been
  correctly aligned.
\item The location of ``export file'' operation has been
  identified. File name has been written to an ImageMagick file.
\item The files produced are equivalent.
\end{enumerate}
The code to perform simple operations is shown in Figure
\ref{fig:IMcode}.

\subsection{Developments to date: R}
\begin{figure}[H]
\includegraphics[width=\columnwidth]{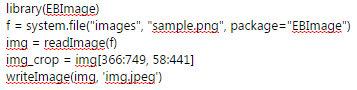}
\caption{EBImage code}
\label{fig:EBcode}
\end{figure}

The ShinyImage tool has been initiated and the basic GUI set up.
The following steps have been taken:
\begin{enumerate}
\item ShinyImage, a prototype image manipulation interface, has been
  developed.  Figure \ref{fig:shinyimage} shows the ShinyImage GUI.
\item The basic representation approach has been determined (use of
  EBImage).
\item The crop, contrast adjustment, and undo/redo operations have
  been implemented. All changed versions of the image are recorded.
\item The user interface is incomplete, and we will be adding pull-down
  menus, online help, more flexible display (e.g. several versions of
  the image displayed side-by-side) and so on.
\item Additional image processing operations will be added.  To
  accommodate larger images, some capability of parallel processing
  will be added.
\item EBImage code is used as the scripting language. The EBImage code
  can either perform the manipulation on the image, or work through
  the GUI. The code is saved in a file during the operation of ShinyImage.
\end{enumerate}
An example of EBImage scripting code is shown in Figure \ref{fig:EBcode}.


\subsection{Plans for the future}
The SWIIM project has made good progress to date. The GIMP track has
set up ImageMagick code to perform the same tasks. The next step is to
devise script-fu code to perform the same processes. Script-fu is more
complex than is ImageMagick. For the R track, the process needs to be
finished to perform the basic tasks on the task list. 

\subsection{Support status}
The SWIIM project has been supported by grants from ORI (ORI2016000141
and ORI2017000232). Support is in place until 2018, with a contingent
extension to 2019.

\section{Conclusions}
Image manipulation is performed interactively in most cases for
scientific images. This is not an optimal situation. The prevalance of
image fraud is high. Image fraud may occur due to ignorance, but it
also seems to occur because images do not agree with the conclusions
that they ``wish'' to draw. As image manipulation is done in a
non-transparent, interactive manner, fraud is common, since some
scientists clearly appear to believe that they can ``get away with
it''.

The SWIIM project intends to improve the manipulation of images, and
to reduce the incidence of image fraud.  The project is modifying
existing image manipulation tools (i.e., GIMP) to produce a
``journal'', a log of the modification process which is also
executable. A journal (an executable log) which performs the same
functions as can be done interactively introduces transparency into
the process. Transparent processes promote honest behavior. In
addition, the journal is executable. Reviewers can, if they wish, run
the journal, and see that the image intended for publication comes
from the source image. This should have the effect of encouraging more
appropriate behavior. In addition, editors of scientific journals can
ask the author of submitted manuscripts to produce these executable
journaled logs.

The R approach is a different, but equally valuable, approach to the
same problem. Many scientists work with R, but do not have a
convenient method to employ scripted image tools. By providing a
method to use the scripted tools in a GUI, but producing a journal log
for later re-use or examination, the R user can extend the ability of
the R system to handle image manipulation.

Scientists can use scripted methods to perform image manipulation with
ImageMagick, EBImage, script-fu, or other scripted tools. This is not
common, however. ImageMagick is not difficult to learn, but there is a
learning curve. Image manipulation is not intuitively done using
scripted commands. Cropping a region (selecting a sub-region in the
image) can be done with a scripted command, but this is not obvious,
because the image is difficult to visualize in terms of
pixels. Interactive methods are much easier to work with.

The journal is a workflow. Thus, the modified image manipulation GIMP
tool is a workflow generator. ShinyImage, a newly constructed image
manipulation tool, performs the image manipulation process with the
same workflow as is saved.  Reproducible AND transparent image
manipulation is possible only with scripts/workflows. By producing the
script/workflow (the journal log) as a consequence of the editing
operation, the best of both worlds is retained. You can have your cake
and eat it too.

\clearpage

\bibliographystyle{ACM-Reference-Format}
\bibliography{../../../reference/repres}

\end{document}